\documentstyle[twocolumn,aps,epsf,floats]{revtex}

\begin{document}

\draft

\twocolumn[\hsize\textwidth\columnwidth\hsize\csname@twocolumnfalse\endcsname

\title {Confinement symmetry, mobility anisotropy, and metallic behavior\\
        in (311)A GaAs 2D holes}
\author{S. J. Papadakis, E. P. De Poortere, and M. Shayegan}
\address{Department of Electrical Engineering, Princeton University,
Princeton, New Jersey  08544, USA.}
\date{\today}
\maketitle
\begin{abstract}
We study two dimensional hole systems confined to GaAs quantum
wells grown on the (311)A surface of GaAs substrates.  Such
samples exhibit an in-plane mobility anisotropy. At constant 2D
hole density, we vary the symmetry of the quantum well potential
and measure the temperature dependence of the resistivity along
two different current directions, in a regime where the samples
exhibit metallic behavior. The symmetry has a significant and
non-trivial effect on the metallic behavior. Moreover, differences
between the temperature-dependence of the resistivity along the
two mobility directions point to specific scattering mechanisms
being important in the expression of the metallic behavior.
\end{abstract}

\vskip1pc]

The unexpected metallic behavior first observed in an Si/SiO$_2$
two-dimensional (2D) electron system \cite{Kravchenko94}, and
subsequently in many other 2D systems
\cite{Popovic97,Coleridge97,Lam97,Hanein98,Simmons98,Papadakis98,Papadakis99,Papadakis00,Murzin98,Yaish99,Mills99}
has generated significant interest.  We study the metallic
behavior in 2D hole systems confined to GaAs quantum wells (QWs)
grown on the (311)A surface of GaAs substrates.  The data we
report demonstrates that both the QW symmetry and the in-plane
mobility anisotropy of (311)A GaAs samples have significant
effects on the low-temperature behavior of the resistivity $\rho$
over a large range of densities.

Hole systems in GaAs QWs have a large spin-orbit interaction,
which in the presence of the inversion asymmetries of the
zincblende crystal structure and of the confining potential,
causes substantial zero-magnetic-field spin-splitting
\cite{Papadakis99,Bychkov2+,Lu98}. Using metal gates placed both
above and below the QW, we tune the symmetry of the QW confinement
potential, and therefore the spin-splitting, while keeping the
density constant \cite{Papadakis99}.  Our goal is to determine the
effect of spin-splitting on the temperature- ($T$) dependence of
$\rho$.

\begin{figure}
\centerline{
\epsfxsize=2.0in
\epsfbox{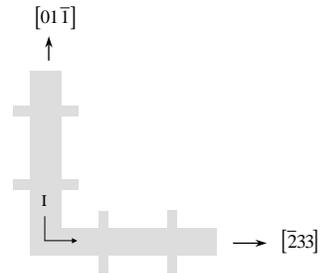}
}
\caption{$L$-shaped Hall bar with which all samples are patterned.
It allows simultaneous measurement of resistivity along the
$[\bar233]$ and $[01\bar1]$ directions.}
\label{LHallbar}
\end{figure}

GaAs (311)A samples also exhibit an in-plane mobility anisotropy.
The anisotropy is caused by irregular corrugations that form at
the interfaces between the GaAs QW and the AlGaAs barriers on
either side \cite{Notzel92,Wassermeier95}. They run parallel to
the $[\bar233]$ direction, causing the mobility along the
$[01\bar1]$ direction to be significantly smaller than the
mobility along the $[\bar233]$ direction. Transport studies
suggest that the scattering mechanisms for current ($I$) along the
two directions have significant differences \cite{Heremans94}: the
mobility of the $[01\bar1]$ direction is limited primarily by
interface-roughness scattering, while the mobility of the
$[\bar233]$ direction is limited by remote ionized impurities. In
order to study the transport properties along both mobility
directions simultaneously, we patterned all of our samples with
the Hall bar of Fig. \ref{LHallbar}.

Our samples are grown by molecular beam epitaxy on undoped (311)A
GaAs substrates, and each consists of a 200 \AA-wide GaAs QW
flanked by AlGaAs barriers which are selectively doped with Si.
They are lithographically patterned with the Hall bar shown
schematically in Fig. \ref{LHallbar} and have metal front and back
gates. Measurements are done in a dilution refrigerator at
temperatures from 0.8 K to 25 mK and in perpendicular magnetic
fields up to 16 T.  We use the low-frequency lock-in technique to
measure the longitudinal ($\rho$) and Hall resistivities.
Experiments are done on samples from four different wafers at 2D
hole densities $p$ from $3.3 \times 10^{11}$ to $2.5 \times
10^{10}$ cm$^{-2}$. These $p$ range from deep in the metallic
regime to near the apparent metal-insulator transition
\cite{Hanein98,Simmons98}. Typical 25 mK mobilities are around 50
m$^2$/Vs.  We concentrate here on data from two samples in the
low-$p$ regime ($p \leq 1.2 \times 10^{11}$ cm$^{-2}$); higher-$p$
data are presented elsewhere \cite{Papadakis99,PapadakisThesis}.

The metal front and back gates are used to alter the symmetry of
the QW, which controls the spin-splitting.  To change the symmetry
at constant $p$, we set the front gate ($V_{fg}$) and back gate
($V_{bg}$) voltages, and measure the Hall resistivity of the
sample. Then, at a small magnetic field, $V_{fg}$ is increased and
the change in $p$ is noted. $V_{bg}$ is then reduced to recover
the original $p$. This procedure applies an electric field
($E_{\perp}$) perpendicular to the plane of the QW while
maintaining $p$ constant (to within 3\% in our experiments).  The
change in $E_{\perp}$ can be calculated accurately from the
dependence of $p$ on the gate voltages. These steps are repeated
until we have probed the range of $V_{fg}$ and $V_{bg}$ that are
accessible without causing gate leakage.  In some samples, this
was done at multiple $p$. While the above method allows accurate
calculation of the change in $E_{\perp}$ when gate voltages are
changed, it does not directly measure the absolute magnitude of
$E_{\perp}$. In the higher $p$ measurements, the absolute
magnitude can be determined directly from the Shubnikov-de Haas
oscillations \cite{Papadakis99}.  In the lower $p$ measurements,
where the sample quality and the magnitude of the spin-splitting
are reduced, we estimate it from the growth parameters of the
samples \cite{PapadakisThesis}, from mobility changes, and from
the magnitude of a magnetoresistance feature at low magnetic field
\cite{Papadakis99,PapadakisThesis}. We believe our estimate to be
accurate to within $\pm1.5$ kV/cm.  In our study, we define
$E_{\perp}$ as positive (negative) if it points to the surface
(substrate).  Here we report data for $E_{\perp} \leq 0$; in cases
where we have been able to explore the $E_{\perp} > 0$ regime
\cite{Papadakis99}, we find the $T$-dependence of $\rho$ to be
nearly symmetric for $E_{\perp} < 0$ and $E_{\perp} > 0$.

Overall, the data show that over a large range of $p$, the
spin-splitting has a significant effect on the $T$-dependence of
$\rho$, and that the specific nature of the different scattering
mechanisms along the two mobility directions is important as well.
This is in contrast to recent suggestions that in this $p$ regime,
knowledge of the mean free path and the Fermi energy is sufficient
to explain the behavior of $\rho$ \cite{Hamilton00}.

\begin{figure}
\centerline{
\epsfxsize=3.22in
\epsfbox{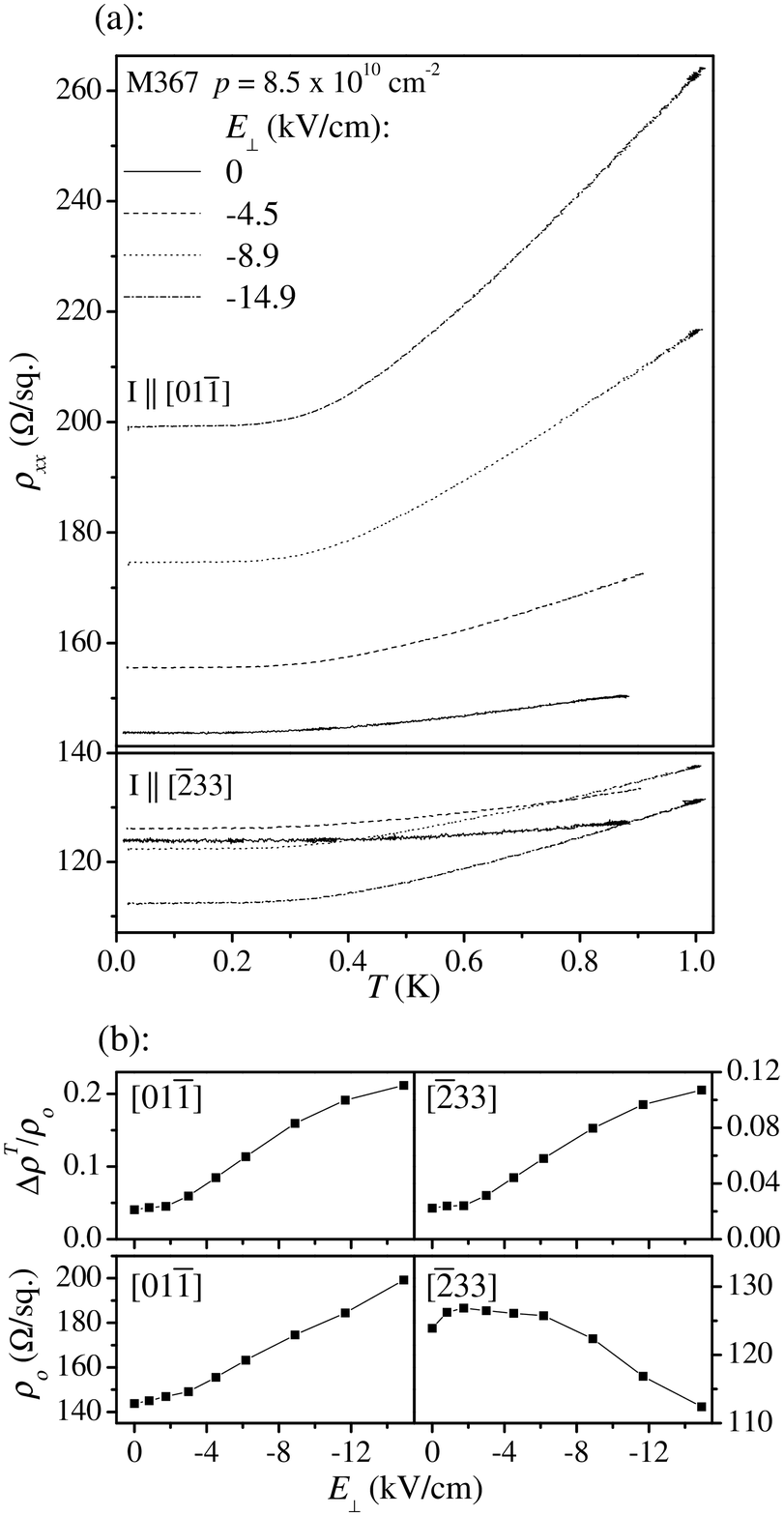}
}
\caption{(a) Selected raw data showing the $T$-dependence of $\rho$ for both current directions.
The top section shows data for $I || [01\bar1]$ and the bottom
section for $I || [\bar233]$.  Both sections are plotted on the
same scale.  (b)  $\Delta\rho^T/\rho_0$ (the fractional change in
$\rho$ from $T$ = 25 mK to 0.8 K) and $\rho_0$ ($\rho$ at $T$ = 25
mK) plotted vs. $E_{\perp}$, shown for the two measured
$I$-directions. The data show that the changes in
$\Delta\rho^T/\rho_0$ as a function of $E_{\perp}$ are not due
simply to changes in $\rho_0$.}
\label{rawdata}
\end{figure}

As an example, we start with data from sample M367 at $p = 8.5
\times 10^{10}$ cm$^{-2}$ (Fig. \ref{rawdata}), after which we
will show that similar effects are observed over a wide range of
$p$.
Concentrating on the top part of Fig. \ref{rawdata}a, we see that
increasing $|E_{\perp}|$ at constant density causes the change in
$\rho$ to become larger as $T$ is increased from 25 mK to 0.8 K
\cite{Papadakis99,Papadakis00}. However, changing $E_{\perp}$ also
affects the resistivity at 25 mK ($\rho_0$), and it has been
suggested that these changes in $\rho_0$ are responsible for the
increased $T$-dependence \cite{Hamilton00}. Looking at data from
both arms of the Hall bar, we are able to determine that it is
mainly the changes in spin-splitting, not in $\rho_0$, that are
causing the changes in the $T$-dependence of $\rho$. The traces in
the top part of Fig. \ref{rawdata}a (above 140 $\Omega$/sq.) are
for the $[01\bar1]$ direction, and those in the bottom part are
for $[\bar233]$. As $|E_{\perp}|$ is increased, $\rho_0$ for the
$[01\bar1]$ direction traces increases, but it decreases for the
$[\bar233]$ direction. However, it is clear that for {\it both}
current directions, as $|E_{\perp}|$ becomes larger the increase
in $\rho$ with $T$ also becomes larger.  Figure \ref{rawdata}b
quantifies these statements.  The upper two panels show the
fractional change in $\rho$, $\Delta\rho^T/\rho_0$, as $T$ is
increased from 25 mK to $\sim 0.8$ K.  Clearly the data from both
$I$ directions have the same behavior as $|E_{\perp}|$ is
increased. The lower two panels show the two $\rho_0$, which do
not have the same behavior as a function of $E_{\perp}$.

\begin{figure*}[tb]
\centerline{
\epsfxsize=5.5in
\epsfbox{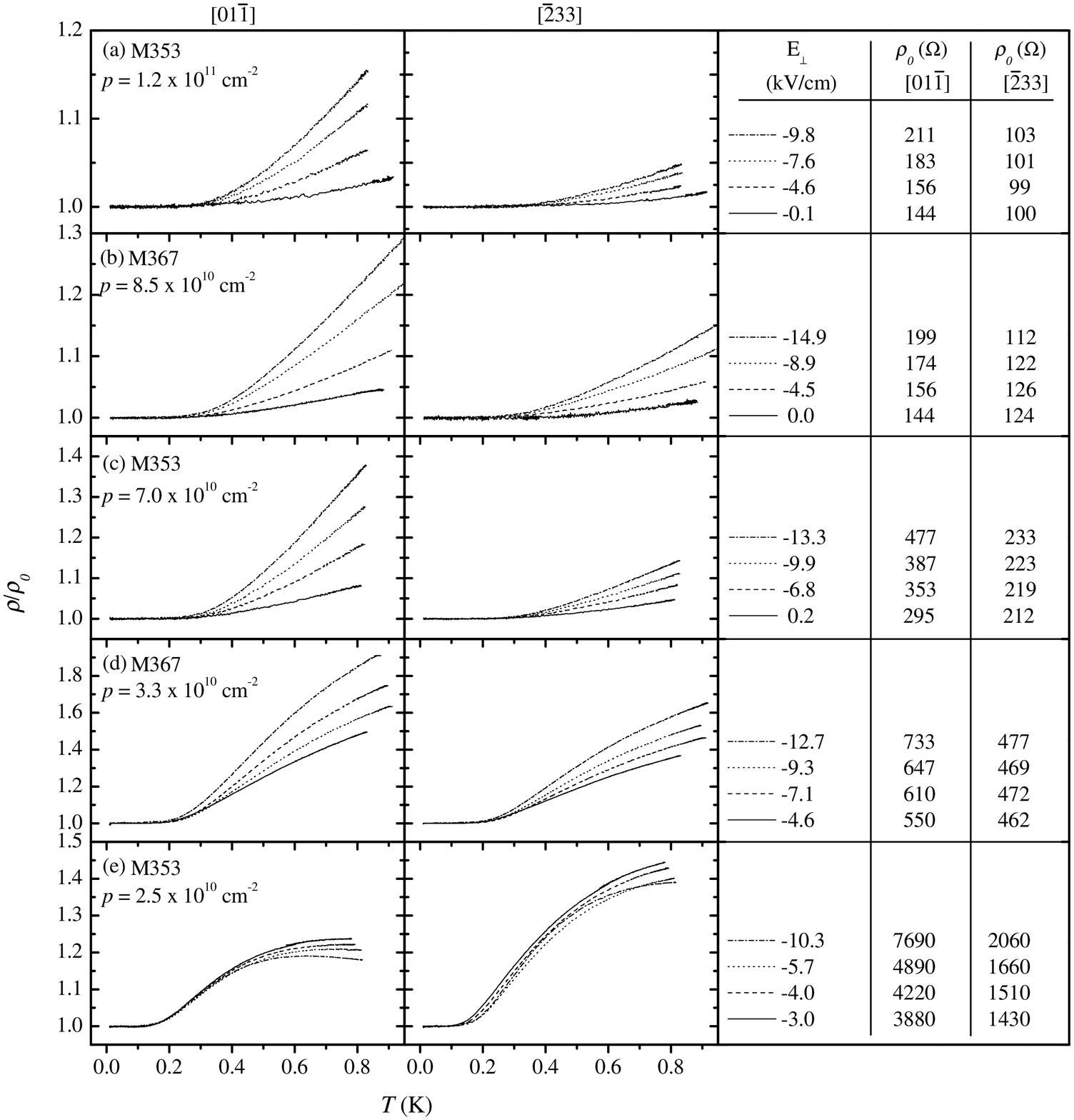}
} \vskip0.5pc
\caption{Fractional change in $\rho$ as a function of $T$,
for a fixed $p$ but varying $E_{\perp}$ in each panel. The $T$ =
25 mK resistivity $\rho_0$ and $E_{\perp}$ for each of the traces
are listed to the right of the figure.}
\label{Tdeps}
\end{figure*}

These data provide strong evidence that the changes in
$\Delta\rho^T/\rho_0$ with $E_{\perp}$ are not driven simply by
the change in the sample disorder as the QW symmetry is varied.
They suggest that the QW symmetry plays a role in the
$T$-dependence of $\rho$ through its effect on the spin-splitting.
We do not mean to imply by this statement that the value of
$\rho_0$ does {\it not} have an effect on the $T$-dependence of
$\rho$.  On the contrary we expect that it does, for in the limit
of a highly disordered sample with very large $\rho_0$ we expect
conventional insulating (strongly localized) behavior.  However,
the data of Fig. \ref{Tdeps} do show that, for small changes in
$\rho_0$ with $E_{\perp}$, the more important parameter is the
symmetry of the QW.

An interesting feature of the data, related to the differences
between the two mobility directions, is that the shapes of the
curves in the top two panels of Fig. \ref{rawdata}b are very
similar, but the magnitudes are different.  For a given
$E_{\perp}$, the spin-splitting is the same regardless of the
$I$-direction in the sample, but the $T$-dependencies of $\rho$
are larger, in {\it both} absolute magnitude and percent change,
for the $[01\bar1]$ direction than for $[\bar233]$. This
emphasizes the subtle nature of the effect of spin-splitting on
the $T$-dependence of $\rho$.  The relationship between the
spin-splitting and the magnitude of the $T$-dependence of $\rho$
is modified differently by the different scattering mechanisms
along the two directions.

We now discuss data from a broad range of $p$. Since plotting the
raw data, as in Fig. \ref{rawdata}a, results in the traces
crossing each other due to their different $\rho_0$, in Fig.
\ref{Tdeps} we plot $\rho/\rho_0$ to make comparison of the traces
more straightforward. As a function of $p$, the data behave
qualitatively as has been observed in previous experiments
\cite{Hanein98,Simmons98}: for the highest $p$ ($p \geq 7.0 \times
10^{10}$ cm$^{-2}$), $\rho$ monotonically increases with
increasing $T$ without inflection, and a reduction in $p$ leads to
a stronger $T$-dependence of $\rho$; for intermediate $p$ ($p =
3.3 \times 10^{10}$ cm$^{-2}$), an inflection point appears in the
measured $T$ range; and for the lowest $p$ ($p = 2.5 \times
10^{10}$ cm$^{-2}$), $\rho$ shows both an inflection point and a
maximum in the measured $T$ range.  If $p$ were decreased further,
we would expect conventional insulating behavior
\cite{Kravchenko94,Hanein98,Simmons98}.

At a given $p$, we can see that the effect on $\rho$ vs. $T$ of an
increase in $|E_{\perp}|$ is qualitatively similar to the effect
of a reduction in $p$.  For the higher $p$ data ($p \geq 7.0
\times 10^{10}$ cm$^{-2}$), increasing $|E_{\perp}|$ increases the
magnitude of the $T$-dependence of $\rho$. In the $p = 3.3 \times
10^{10}$ cm$^{-2}$ data, increasing $|E_{\perp}|$ both strengthens
the $T$-dependence of $\rho$ and moves the inflection point to
lower $T$. In the $p = 2.5 \times 10^{10}$ cm$^{-2}$ data,
increasing $|E_{\perp}|$ causes both the inflection point and the
maximum to shift to lower $T$ and leads to {\it weaker} overall
$T$-dependence of $\rho$ (again qualitatively similar to the
effect of reducing $p$ in this regime).  An interesting point is
that there is likely a $p$ where $\rho$ may show very similar
fractional change with $T$ as $E_{\perp}$ is varied.  There is
another possible reason for the different behavior in the lowest
$p$ data: at $p = 2.5 \times 10^{10}$ cm$^{-2}$, $\rho_0$
increases significantly more as a function of $E_{\perp}$ than it
does at the other $p$.

As was done in the data at $p = 8.5 \times 10^{10}$ cm$^{-2}$
(Fig. \ref{rawdata}), at the other $p$ differences in the
behaviors of $\rho_0$ vs. $E_{\perp}$ from the two arms of the
Hall bar can be used to separate the effects of $E_{\perp}$ and
$\rho_0$. Along the $[01\bar1]$ direction, it is typical for
$\rho_0$ to increase as $E_{\perp}$ is increased. This is because
interface roughness scattering is the limiting factor for the
mobility \cite{Heremans94}.  As the QW is made more asymmetric,
the hole wavefunction is pushed towards one edge of the QW, so it
is more affected by the interface roughness, and $\rho_0$
increases. Along the $[\bar233]$ direction, however, the interface
roughness plays a much smaller role, so $\rho_0$ is limited by
ionized impurity scattering. While the behavior of $\rho_0$ along
$[01\bar1]$ as a function of $E_{\perp}$ in all samples is
qualitatively similar to that in Fig. \ref{rawdata}b, the behavior
of $\rho_0$ along $[\bar233]$ varies significantly with sample and
$p$. For example, it is non-monotonic in sample M353 at $p = 1.2
\times 10^{11}$ cm$^{-2}$, and in M367 at $p = 8.5$ and $3.3
\times 10^{10}$ cm$^{-2}$ \cite{rho0acc}. In all samples at all
measured $p$, $\Delta\rho^T/\rho_0$ has the same qualitative
dependence on $E_{\perp}$ in both $I$ directions.

The data presented in this report, along with those reported in
Refs. \cite{Papadakis99} and \cite{PapadakisThesis}, show that for
seven densities, measured in samples from four different wafers,
covering more than an order of magnitude of $p$ ($2.5 \times
10^{10}$ to $3.3 \times 10^{11}$ cm$^{-2}$), the asymmetry of the
QW, and therefore the magnitude of the spin-splitting, has an
effect on the low-$T$ temperature-dependence of $\rho$. In the
range $p \geq 3.3 \times 10^{10}$ cm$^{-2}$, as $p$ is reduced the
magnitude of the $T$-dependence of $\rho$ becomes larger.
Interestingly, in this range the effect of $E_{\perp}$ also
becomes larger, in that the same $E_{\perp}$ causes a larger
increase in $\rho/\rho_0$ with increasing $T$. This is especially
noteworthy because the spin-splitting is becoming smaller as $p$
is reduced, so the effect of spin-splitting on the metallic
behavior is becoming much stronger as $p$ is reduced from $3.3
\times 10^{11}$ to $3.3 \times 10^{10}$ cm$^{-2}$
\cite{Papadakis00c}. Note also that the data of Fig. \ref{Tdeps}
imply that the effect of $E_{\perp}$ is strongest at $p$ where the
metallic behavior is most pronounced.  This emphasizes the link
between the spin-splitting and the metallic behavior in GaAs 2D
holes \cite{Papadakis99,Murzin98,Yaish99,Papadakis00b}.

In summary our data reveal two specific parameters which affect
the $T$-dependence of $\rho$ significantly. First, the symmetry of
the QW plays a role through spin-splitting in both the magnitude
of the change in $\rho$ with $T$ and in the shape of the curve.
Second, the absolute and fractional changes in $\rho$ are
quantitatively different for the two mobility directions, pointing
out that interface roughness scattering also has an important
role. A model including only ionized impurity scattering has
reproduced some of the qualitative features of $\rho$ vs. $T$
\cite{DasSarma99}. We propose that the $T$-dependencies of both
inter-spin-subband scattering and interface roughness scattering
should also be included in calculations which attempt to explain
$\rho$ vs. $T$ data.  The different dependencies of $\rho$ on the
various parameters reveal that the metallic behavior cannot be
simply characterized by any overall parameter such as mean free
path or Fermi energy \cite{Feng+}.

We gratefully acknowledge the NSF and DOE for supporting this
work.


\begin{thebibliography}{10}
\bibitem{rho0acc}
Note that the measured $\rho_0$ are very accurate as they result
from averaging
  $\rho$ at base-$T$ for a long period of time.
\end{thebibliography}

\end{document}